\documentclass[12pt,preprint]{elsarticle}
\usepackage{graphicx}
\usepackage{amsmath,amssymb}
\usepackage{lineno}
\journal{Phys.\,Lett.\,B (13 Aug.\,2019)}

\usepackage{texnames}
\usepackage[T1]{fontenc}
\usepackage{hepnames}
\usepackage{siunitx}
\usepackage{color}
\usepackage{nicefrac}
\usepackage{hyperref}
\hypersetup{colorlinks = true, allcolors = blue}
\usepackage{bold-extra}
\usepackage{soul}

\begin{document}

\begin{frontmatter}

\title{Beam-beam effects on the luminosity measurement at LEP and the number of light neutrino species\\
{\small \it Marking the 30th anniversary of the first Z detected at LEP on 13 August 1989}}

\author[1]{Georgios Voutsinas}
\ead{georgios.gerasimos.voutsinas@cern.ch}
\author[1]{Emmanuel Perez}
\ead{Emmanuel.Perez@cern.ch}
\author[2]{Mogens Dam}
\ead{dam@nbi.dk}
\author[1]{Patrick Janot}
\ead{Patrick.Janot@cern.ch}

\address[1]{CERN, EP Department, 1 Esplanade des Particules, CH-1217 Meyrin, Switzerland} 
\address[2]{Niels Bohr Institute, University of Copenhagen, Blegdamsvej 17,
  2100 Copenhagen, Denmark}


\begin{abstract}
{\small 
In $\rm e^+ \rm e^-$ collisions, electromagnetic effects caused by large charge density bunches modify the effective acceptance of the luminometer system of the experiments.
These effects consequently bias the luminosity measurement from the rate of low-angle  Bhabha interactions $\rm e^+ e^- \to e^+ e^- $.
Surprisingly enough, the magnitude of this bias is found to yield an underestimation of the integrated luminosity measured by the LEP experiments by about 0.1\%,  significantly larger than the reported experimental uncertainties. When accounted for, this effect modifies the number of light neutrino species determined at LEP from the measurement of the hadronic cross section at the Z peak.} 

\end{abstract}


\date{Revised 14 Sept.\,2019}

\end{frontmatter}

\vfill\eject

\section{Introduction}

The Large Electron-Positron (LEP) collider was operated at CERN between $1989$ and $2000$, and delivered $\rm e^+ e^-$ collisions to four experiments, at centre-of-mass energies that  covered the $\rm Z$ resonance, the $\rm WW$ threshold, and extended up to $\sqrt{s} = 209$\,GeV. The first phase (LEP1), at and around the $\rm Z$ pole, provided a wealth of  measurements of unprecedented accuracy~\cite{ALEPH:2005ab}. In particular, the measurement of the hadronic cross section at the Z peak, $\sigma_{\rm had}^0$, has been used to derive the number of light neutrino species $N_{\nu}$ from
\begin{equation}
N_\nu \left( \frac{\Gamma_{\nu\nu}}{\Gamma_{\ell\ell}}\right)_{\rm SM}= \left( \frac{12\pi}{m_{\rm Z}^2} \frac{R^0_\ell}{\sigma^0_{\rm had}} \right)^{\frac{1}{2}} - R^0_\ell - 3 - \delta_\tau, 
\label{eq:NnuForLep}
\end{equation}
where $R^0_\ell$ is the ratio of the hadronic-to-leptonic Z branching fractions; $\delta_\tau$ is a small ${\cal{O}}(m_\tau^2/m_{\rm Z}^2) $ correction; and $(\Gamma_{\nu\nu}/\Gamma_{\ell\ell})_{\rm SM}$ is the ratio of the massless neutral-to-charged leptonic Z partial widths predicted by the Standard Model (SM). The combination of the measurements made by the four LEP experiments leads to~\cite{ALEPH:2005ab}:
\begin{equation}
   N_{\nu} = 2.9840 \pm 0.0082,
   \label{eq:Nnu}
\end{equation}
consistent within two standard deviations with the three observed families of fundamental fermions.\footnote{The expression in Eq.~\ref{eq:NnuForLep} was chosen to minimize the dependence of $N_{\nu}$ on SM parameters. With up-to-date calculations of higher-order corrections to $(\Gamma_{\nu\nu}/\Gamma_{\ell\ell})_{\rm SM}$~\cite{Tanabashi:2018oca} and recent measurements of the Higgs boson and top quark masses, the number of light neutrino species slightly increases from 2.9840 to 2.9846. The more parameter-dependent global fit of Ref.~\cite{Tanabashi:2018oca}, which also includes the Z width measurement and the world-average value of the strong coupling constant, yields $N_\nu = 2.991\pm 0.007$, with a similar sensitivity to the LEP integrated luminosity~\cite{Jens}.} This observable is directly affected by any systematic bias on the integrated luminosity through $\sigma^0_{\rm had}$. Indeed, the integrated lumino\-sity uncertainty saturates the uncertainty on $\sigma^0_{\rm had}$, and is the largest contribution to the $N_{\nu}$ uncertainty.

At LEP, the luminosity was determined by measuring the rate of the theoretically well-understood Bhabha-scattering process at small angles, $\rm e^+ e^- \to \rm e^+ \rm e^- $, in a set of dedicated calorimeters (LumiCal), possibly completed with tracking devices,  situated on each side of the interaction region. These luminometers covered polar angle ranges from about $25$ to $60$\,mrad ($29$ to $185$\,mrad for DELPHI) from the beam axis. The Bhabha events were selected with a "narrow" acceptance on one side and a "wide" acceptance on the other, defined as shown in Table~\ref{tab:AngularRanges}.

\begin{table}[htbp]
\begin{center}
\vspace{-3mm}
\caption{Wide and narrow acceptance for the second-generation LumiCals of the four LEP experiments between 1993 and 1995 (1994--95 for DELPHI). %
\vspace{2mm} }
\label{tab:AngularRanges}
\begin{tabular}{|l|c|c|c|c|}
    \hline
    Experiment     &   ALEPH~\cite{Buskulic:1994wz} & DELPHI~\cite{Camporesi:2629467} & L3~\cite{Brock:307585}
  &  OPAL~\cite{Abbiendi:1999zx}  \\ \hline\hline
    Wide (mrad) & 26.2--55.5 & 37.0--127.0 & 27.0--65.0 & 27.2--55.7 \\ \hline
    Narrow (mrad) & 30.4--49.5 & 44.9--113.6 & 32.0--54.0 & 31.3--51.6 \\ \hline
\end{tabular}
\end{center}
\end{table}

\begin{figure}[htb]
 \centering
\includegraphics[width=0.85\columnwidth]{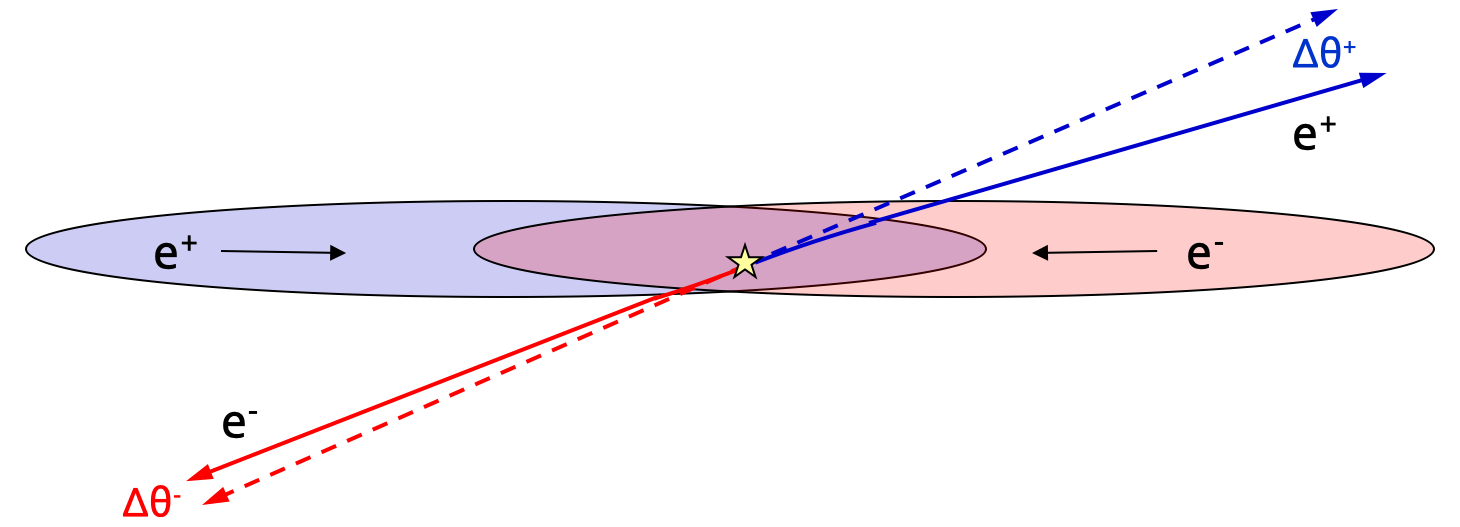} 
 \caption{Illustration of the effect of the focusing Lorentz force experienced by the charged leptons emerging from a Bhabha interaction. The dashed lines show the original direction of the leptons, while the full lines show their direction after the electromagnetic deflection induced by the opposite charge bunch.
 }
 \label{fig:sketcha}
\end{figure}

When the charge density of the beam bunches is large, beam-induced effects modify the effective acceptance of the LumiCal in a nontrivial way. The final state $\rm e^+$ ($\rm e^-$) in a Bhabha interaction, emitted at a small angle off the $\rm e^+$ ($\rm e^-$) beam, feels an attractive force from the incoming $\rm e^-$ ($\rm e^+$) bunch, and is consequently focused towards the beam axis.\footnote{The ``repelling'' effect of the particle's own bunch is negligible because, in the laboratory frame, the electric and magnetic components of the Lorentz force have the same magnitude but opposite directions. In contrast, the electric and magnetic forces induced by the opposite charge beam point in the same direction and thus add up.} This effect, illustrated in Fig.~\ref{fig:sketcha}, 
leads to an effective reduction of the acceptance of the LumiCal, as particles that would otherwise hit the detector close to its inner edge are focused to lower polar angles and may therefore miss the detector.

This effect has been first realised in the context of the International Linear Collider (ILC) design study~\cite{Rimbault:2007zz}.
A detailed analysis of beam-induced effects on the luminosity measurement at the Future Circular Collider (FCC) has been recently carried out and is reported in Ref.~\cite{Voutsinas2019}, together with methods to measure and correct for them. In the context of that study, it has been realised that these effects were already significant
at LEP. As shown below, they lead to a bias  of the measured luminosity  of the order of 
$0.1 \%$, which is large compared to the uncertainties reported by the experiments.\footnote{The most precise determination, from the OPAL experiment~\cite{Abbiendi:1999zx}, quotes an experimental uncertainty of $0.034 \%$ and a theoretical uncertainty of $0.054 \%$. 
} 
Needless to say, beam-beam effects were already well-known at the time of LEP~\cite{Brandt:2000xk}.
To our knowledge, however, this beam-induced bias of the luminosity measurement has not been
taken into account by the  LEP experiments.

In the study presented here, this bias is quantified with the {\tt Guinea-Pig} code~\cite{Schulte:1999tx}. In Section~\ref{sec:NumericalCalculations}, technical details are given on how this code is used to perform the numerical calculations. Detailed results of these calculations are presented in Section~\ref{sec:focusing}, for a representative situation corresponding to the OPAL luminosity measurement performed in  $1994$, during which half of the LEP data at the $\rm Z$ peak was collected. This illustrative example assumes a polar angle acceptance between $\theta_{\rm{min}} = 31.3$\,mrad and $\theta_{\rm{max}} = 51.6$\,mrad (Table~\ref{tab:AngularRanges}), and the set of beam parameters given in the third row of Table~\ref{tab:parameters}. 
\begin{table}[htbp]
\begin{center}
\caption{Parameters for the LEP operation at the Z pole in 1993, 1994, and 1995, relevant for the determination of the beam-induced luminosity bias: number of particles per bunch ($N$), horizontal ($\sigma_x$) and vertical ($\sigma_y$) bunch sizes, longitudinal bunch length ($\sigma_z$), and values of the $\beta$ function at the interaction point in the $x$ and $y$ directions. The number of particles per bunch and the bunch sizes are (instantaneous-luminosity-weighted) averaged over the year, as described in Section~\ref{sec:Systematic}.\vspace{2mm}}
\label{tab:parameters}
\begin{tabular}{|c|c|c|c|c|c|c|}
    \hline
            & $N$            & $\sigma_x$  & $\sigma_y$   &   $\sigma_z$  &   $\beta^*_x$  & $\beta^*_y$  \\
    Year    & ( $10^{11}$ )  & ( $\mu$m )  &  ( $\mu$m )  &  ( mm )       &  ( m )  &  ( cm )   \\ \hline\hline
    1993    &  1.207          &      213.   &   $\sim 4.$  &   10.3         &   2.5   &    5.     \\ \hline
    1994    &  1.280          &      171.   &   $\sim 4.$  &   10.0         &   2.0  &    5.     \\ \hline
    1995    &  1.155          &      206.   &   $\sim 4.$  &   10.5         &   2.5   &    5.     \\ \hline
\end{tabular}
\end{center}
\end{table}

The calculation is extended in Section~\ref{sec:neutrinos} to the four LEP experiments and to the last three years of LEP1 operation (1993, 1994, and 1995), when the experimental precision of the integrated luminosity measurements was improved by up to one order of magnitude with the installation of second-generation LumiCals. A corrected number of light neutrino species from the combination of the LEP measurements is deduced. In Section~\ref{sec:Systematic}, the systematic effects arising from the simplifying assumptions used to determine $N_\nu$ are evaluated and corrected for. A summary is given in Section~\ref{sec:summary}.


\section{Numerical Calculations}
\label{sec:NumericalCalculations}

The {\tt Guinea-Pig} code~\cite{Schulte:1999tx} was initially developed in the mid-nineties to simulate the beam-beam effects and the beam-background production in the interaction region of (future) electron-positron colliders. 
The {\tt Guinea-Pig} algorithm groups particles from the incoming bunches into macro-particles, slices each beam longitudinally, and divides the transverse plane into a ``grid'' of cells. The macro-particles are initially distributed over the slices and the grid, and are tracked through the collision. The fields are computed at the grid points at each step of this tracking. 
Here, the dimensions of the grid are defined to contain the $\pm 3 \sigma_z$ envelope of the beam in the longitudinal direction, and the $\pm 3 \sigma_x$ and $\pm 6 \sigma_y$ intervals in the transverse dimensions. The number of cells (slices) are such that the cell (slice) size, in both the $x$ and $y$ dimensions (along the $z$ axis), amounts to about $10 \%$ of the transverse (longitudinal) bunch size at the interaction point.

In the context of the studies reported in Ref.~\cite{Rimbault:2007zz}, the C++ version of {\tt Guinea-Pig} was extended in order to track Bhabha events, provided by external generators, 
in the field of the colliding bunches. This version of {\tt Guinea-Pig} is used here. An input Bhabha event is associated to one of the $\rm e^+ \rm e^-$ interactions, i.e., is assigned a spatial vertex and an interaction time according to their probability densities. 
The electron and positron that emerge from this Bhabha interaction 
are subsequently transported as they move forward: the final state $\rm e^-$ ($\rm e^+$) potentially crosses a significant part of the $\rm e^+$ ($\rm e^-$) bunch, or travels for some time in its vicinity and, thereby, feels a deflection force.

Since the $\rm e^{\pm}$ that emerge from a Bhabha interaction are emitted with a non-vanishing, albeit small, polar angle, they may exit the grid mentioned above, designed to contain the beams  and in which the fields are computed, before the tracking ends. For this reason, the program can also extend the calculations of the fields to ``extra'' grids. For the settings used here, six extra grids are defined to sample larger and larger spatial volumes with accordingly decreasing granularity. 
The largest grid has equal dimensions in $x$ and $y$, and a size twelve times larger than that of the first grid in $x$. It safely contains the trajectory of Bhabha electrons during the whole tracking time for the range in polar angle of interest here. 

A numerical integration code has also been developed, which uses the Bassetti-Erskine formulae~\cite{Bassetti:1980by} for the field created by a Gaussian bunch to determine the average effects that a particle would feel. 
The particle is defined by its velocity and spatial coordinates at a given time $t_0$. The momentum kick that it gets between $t_0$ and a later time is obtained by integrating the Lorentz force during this interval~\cite{Keil:1994dk}. More details are given in Ref.~\cite{Voutsinas2019}.


\section{Electromagnetic focusing of final state leptons in Bhabha events} 
\label{sec:focusing}

The {\tt Guinea-Pig} code is used in this section to estimate the focusing of final-states leptons, first for {\it leading-order} Bhabha events (i.e., without initial- or final-state radiation), with the 1994 LEP beam parameters given in Table~\ref{tab:parameters}. The corresponding luminosity bias is evaluated for the illustrative example of the OPAL LumiCal narrow acceptance, $31.3 < \theta < 51.6$\,mrad. In what follows, the polar angle $\theta$ of the electron (positron) emerging from a Bhabha interaction is always defined with respect to the direction of the $\rm e^-$ ($\rm e^+$) beam. The notation $\theta^*$ is used to denote the production angles in the frame where the initial $\rm e^+ \rm e^-$ pair is at rest, while $\theta^0$ labels these angles in the laboratory frame. The {\tt Guinea-Pig} simulation includes the intrinsic transverse momentum  of the particles in the bunches, so that the initial $\rm e^+ \rm e^-$ pair is not strictly at rest in the laboratory frame, which in turn creates a smearing by a few tens of $\mu$rad around the $\theta^*$ production angle. The mean of $\theta^0 - \theta^*$ is zero irrespective of the kinematic properties of the event.

\begin{figure}[htbp]
\begin{center}
\begin{tabular}{cc}
\includegraphics[width=0.48\columnwidth]{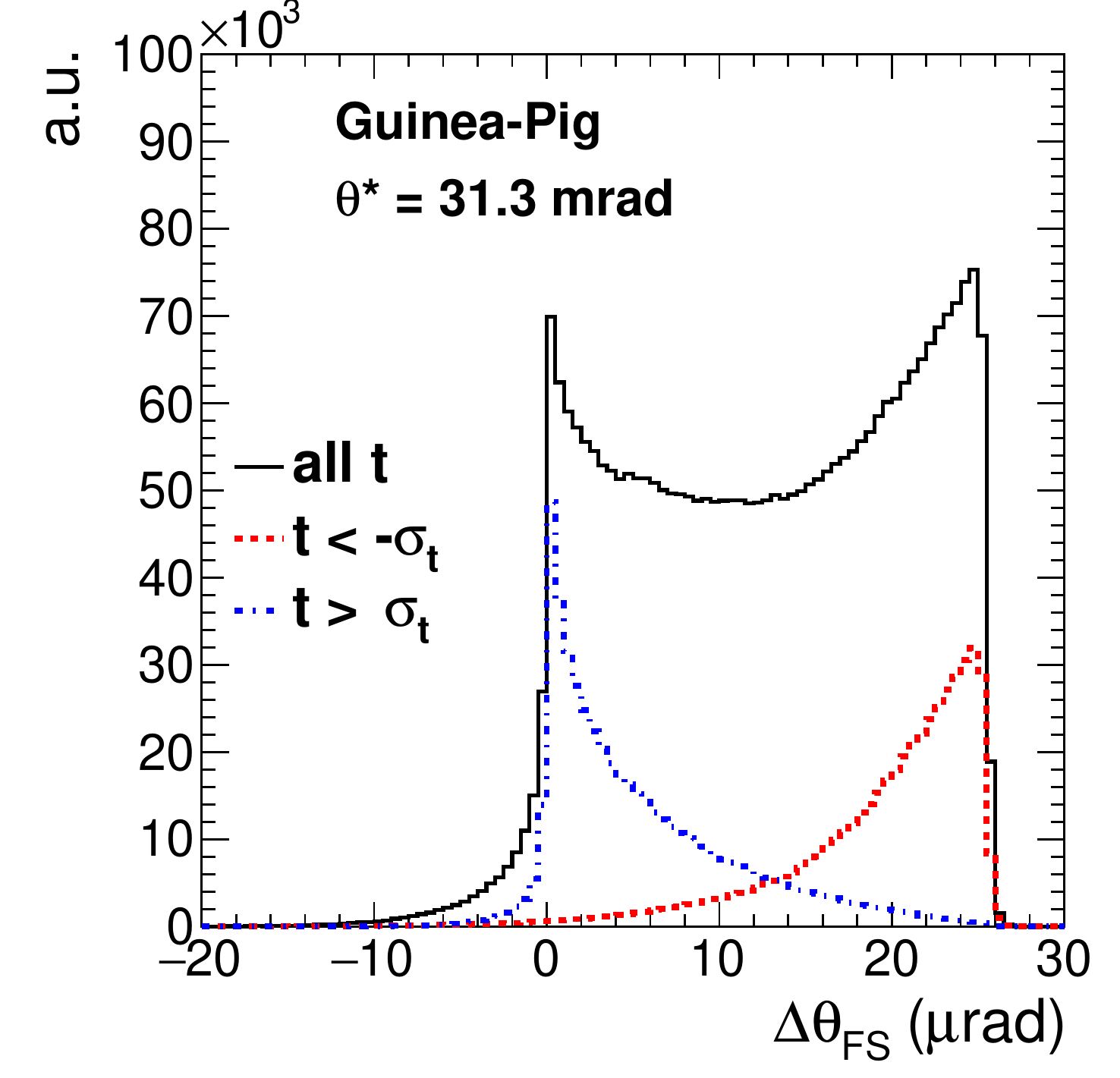} &
\includegraphics[width=0.48\columnwidth]{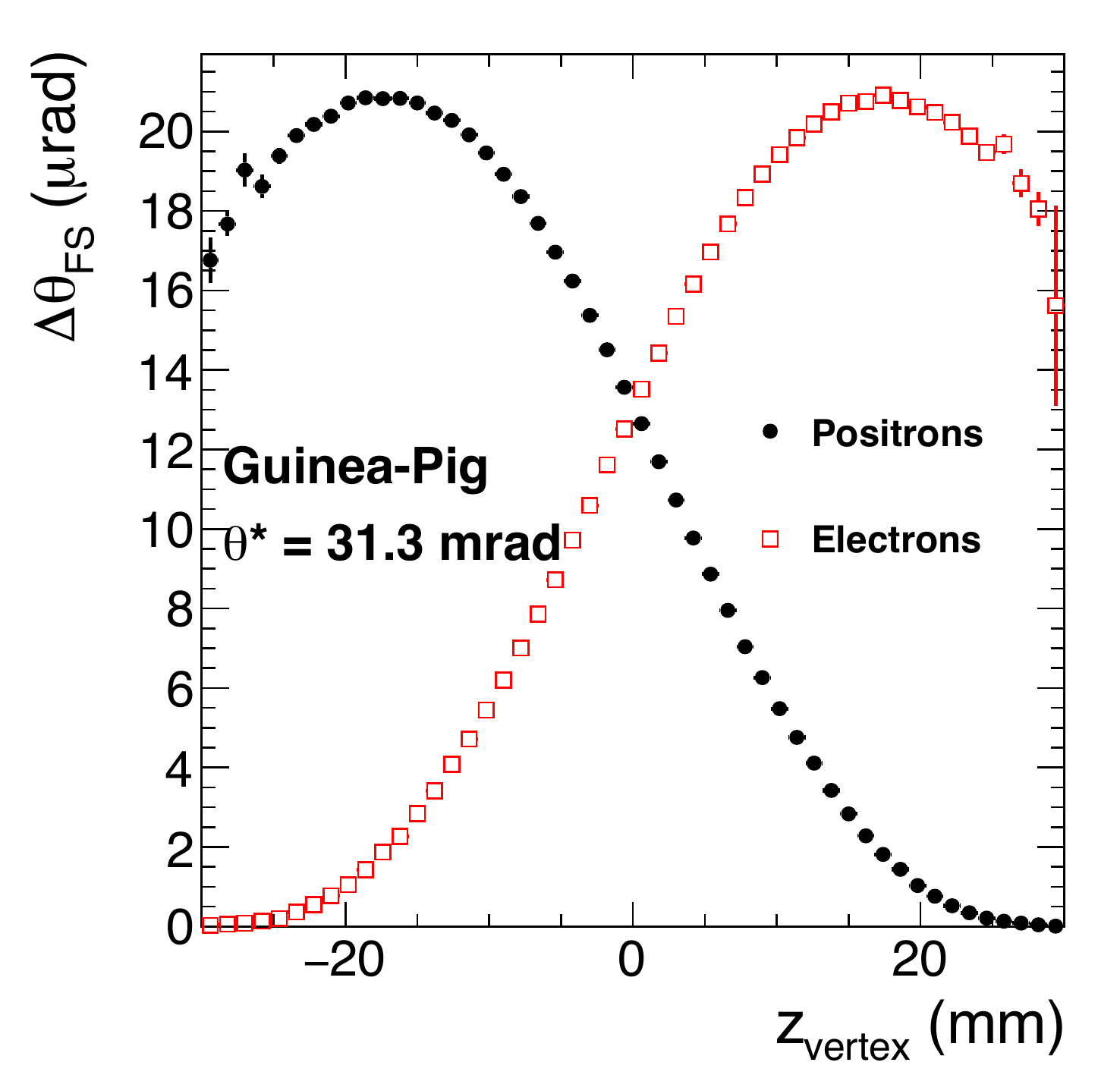}
\end{tabular}
\caption{\small Left: Distribution of the angular focusing $\Delta \theta_{\rm{FS}}$ for $45.6$\,GeV electrons produced at $\theta^* = 31.3$\,mrad, as predicted by {\tt Guinea-Pig}, for (full black line) all events, (dash-dotted blue line) events corresponding to ``late'' interactions and (dashed red line) events corresponding to ``early'' interactions. The latter (former) interactions occur by definition at a time $t < -\sigma_t$ ( $t > \sigma_t$), with $\sigma_t = \sigma_z / \sqrt{2}c$, the  origin being given by the time when the centres of the two bunches coincide. Right: Average deflection of $45.6$\,GeV leptons emerging from a Bhabha interaction at $\theta^* = 31.3$\,mrad, as a function of the longitudinal position of the vertex of the interaction, shown separately for the ${\rm e^-}$ (open squares) and the ${\rm e^+}$ (closed dots). The ${\rm e^+}$ beam moves towards the positive $z$ direction.
}
\label{fig:deltaTheta}
\end{center}
\end{figure}

As depicted in Fig.~\ref{fig:sketcha}, the electrons and positrons emerging from a Bhabha interaction experience the field of the opposite charge bunch.  The left panel of Fig.~\ref{fig:deltaTheta} shows the distribution of the angular deflection $\Delta \theta_{\rm{FS}}$ of $45.6$\,GeV electrons emitted at
a fixed angle $\theta^\ast = 31.3$\,mrad, as predicted by {\tt Guinea-Pig}. 
It is defined as the difference between the polar angle of the outgoing electron before and after this deflection, $\Delta \theta_{\rm{FS}} = \theta_{0} - \theta$, where $\theta$ denotes the final polar angle,  such that a positive quantity corresponds to a focusing deflection towards the beam direction.

For ``late'' interactions that occur after the cores of the two bunches have crossed each other, the final state $\rm e^{\pm}$ do not see much of the $\rm e^{\mp}$ bunch charge and many of them are minimally deflected. On the contrary, for ``early'' interactions that take place when the two bunches just start to overlap, the emitted $\rm e^{\pm}$ travel through the whole $\rm e^{\mp}$ bunch and are largely deflected. These observations explain the two peaks seen in Fig.~\ref{fig:deltaTheta}.

The right panel of Fig.~\ref{fig:deltaTheta} shows the average deflection of such electrons as a function of the interaction vertex position $z_{\rm{vtx}}$. Here, the positrons have a positive momentum in the $z$ direction, such that, when $z_{\rm{vtx}}$ is large and negative, they cross the whole electron bunch. In contrast, they see little charge from this bunch when $z_{\rm{vtx}}$ is large and positive, resulting in a vanishing deflection. The electron deflection follows a symmetric behaviour. 

\begin{figure}[htbp]
\begin{center}
\begin{tabular}{cc}
\includegraphics[width=0.48\columnwidth]{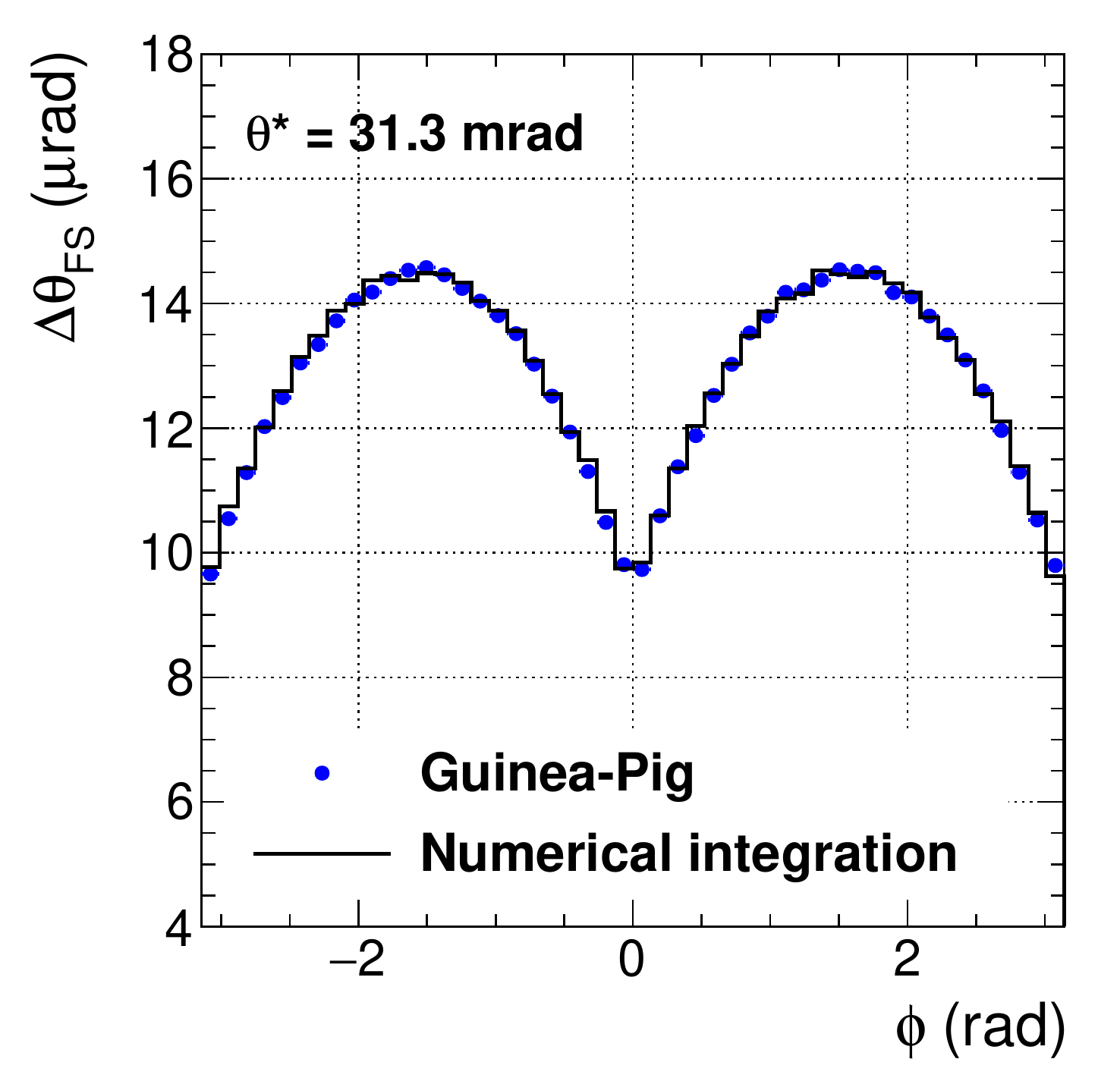} & 
\includegraphics[width=0.48\columnwidth]{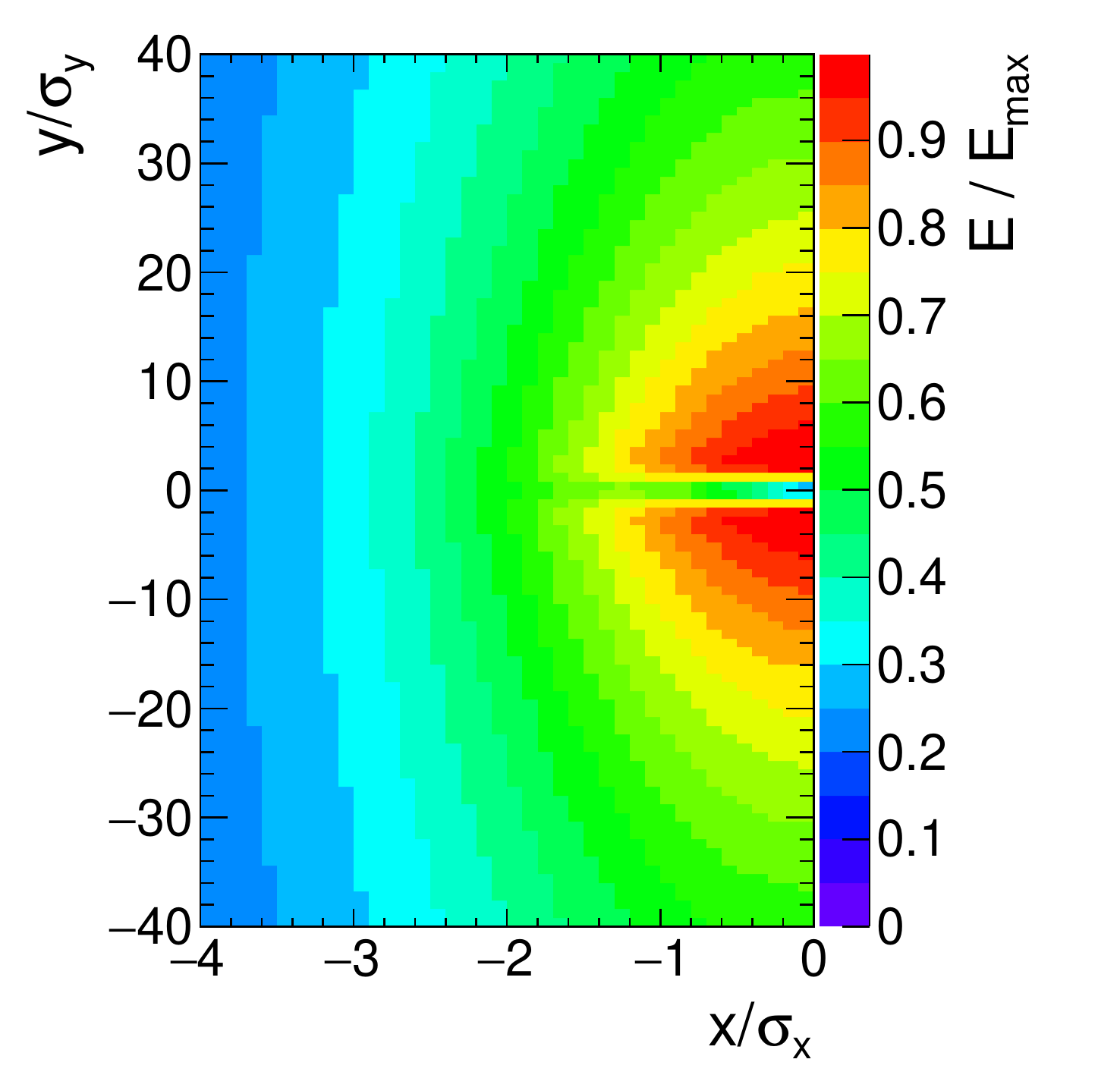} 
\end{tabular}
\caption{\small 
Left: $\Delta \theta_{\rm{FS}}$ for $45.6$\,GeV electrons produced at an angle $\theta^* = 31.3$\,mrad, as a function of their azimuthal angle, as predicted by {\tt Guinea-Pig} and by a numerical integration of the average Lorentz force felt by the electrons. 
Right: Electric field strength $E$ created by a bunch in the laboratory frame, shown as a function of the $(x,y)$ coordinates in any transverse cross section, and normalized to the maximum field strength $E_{\rm max}$ in this cross section.}
\label{fig:dtheta_vs_phi_and_field}
\end{center}
\end{figure}

The left panel of Fig.~\ref{fig:dtheta_vs_phi_and_field} shows
how the strength of the focusing strongly depends on the azimuthal angle $\phi$ of the electrons: it is maximal for electrons emitted vertically at $\phi = \pm \pi/2$ and smaller by about $30 \%$ for electrons emerging horizontally at $\phi =0$ or $\phi = \pi$. This plot also shows that the {\tt Guinea-Pig} simulation and the numerical integration mentioned in Section~\ref{sec:NumericalCalculations} are in agreement. The $\phi$ dependence reflects the fact that, since the bunches are flat with $\sigma_y \ll \sigma_x$, the electromagnetic field created by the bunches is much stronger along the $y$ than along the $x$ direction, as illustrated in the right panel of Fig.~\ref{fig:dtheta_vs_phi_and_field}. 

For electrons emerging close to the lower (upper) edge of the narrow LumiCal acceptance, the average deflection amounts to $12.81\,\mu$rad ($11.19\,\mu$rad). The net effect is that the number of electrons detected in the LumiCal, in the range  $\theta_{\rm{min}} < \theta < \theta_{\rm{max}}$, is smaller than the number of Bhabha electrons emitted within this range, which leads to an underestimation of the luminosity. From the expression of the counting rate of leading-order Bhabha events in the LumiCal,

$$ N \propto \int_{ \theta_{\rm min}}^{\theta_{\rm max}} \frac{ d \theta} { \theta^3}  ,$$ 
the bias induced by this angular deflection reads

\begin{equation}
 \frac{\Delta N}{N} = \frac{ -2 } { \theta_{\rm{min}}^{-2} - \theta_{\rm{max}}^{-2} } \left( \frac{ \Delta \theta_{\rm{FS}} (\theta = \theta_{\rm{min}})}{\theta_{\rm{min}}^3} - \frac{ \Delta \theta_{\rm{FS}} (\theta = \theta_{\rm{max}})}{\theta_{\rm{max}}^3} \right )  ,
 \label{eq:lumiBias}
 \end{equation}
which, numerically, leads to a bias of the measured luminosity by $\Delta L / L \simeq -0.1059\%$.
This effect is larger than the experimental uncertainty of the luminosity measurement reported in Ref.~\cite{Abbiendi:1999zx} for 1994. \\

To assess the effect of higher-order electroweak corrections (including initial-state and final-state radiation, ISR and FSR), a sample of about four million Bhabha events, produced with the {\tt BHLUMI~4.04} Monte-Carlo event generator~\cite{Jadach:1996is}, from which the LEP experiments determined their acceptance, was used to estimate a multiplicative ``$k$ factor'' on the luminosity bias. Both ISR and FSR lead to softer electrons and positrons in the final state, which thus experience a stronger focusing. This effect tends to increase the luminosity bias by typically 5\%. In addition, ISR causes a longitudinal boost of the final-state leptons, resulting in a nontrivial effect on the effective LumiCal acceptance and therefore on the luminosity bias, which gets reduced by typically 1.5\%. Finally, FSR photons are usually emitted at a small angle with respect to the final state leptons, and smear their direction symmetrically at larger and smaller angles, yielding a second-order effect on average. A clustering algorithm is likely to merge the electron and the radiated photon into a single cluster, thereby compensating for this small effect. An accurate evaluation of the latter, which would probably require the {\tt BHLUMI} events to be processed through a full simulation of the LumiCal and a cluster reconstruction algorithm to be run on the simulated energy deposits, is therefore not crucial. 

The ISR/FSR-corrected luminosity bias is determined here with {\tt BHLUMI} events from the kinematic properties of the final-state charged leptons only. With a loose lower-energy cut of $5$\,GeV on the final-state leptons, a bias of $-0.1126\%$ is obtained. If it is required in addition, as in Ref.~\cite{Abbiendi:1999zx}, that both leptons have an energy above $E_{\rm{beam}}/2$; that the average of their energies exceed $0.75 \times E_{\rm{beam}}$, with $E_{\rm{beam}} = \sqrt{s}/2$; and that their acoplanarity $\vert \vert \phi^+ - \phi^- \vert - \pi \vert$  and acollinearity $\vert \theta^+-\theta^- \vert$ be smaller than 200 and 10\,mrad, respectively; the resulting bias amounts to $-0.1113\%$, corresponding to a $k$ factor of $1.051$ with respect to leading-order Bhabhas.

\section{Luminosity bias and impact on the measurement of the number of light neutrino species} 
\label{sec:neutrinos}

The calculation presented in the previous section for OPAL in 1994 was repeated for the four experiments and the last three years of LEP1 operation (1993--1995). The corresponding angular ranges are shown in Table~\ref{tab:AngularRanges}, and the LEP beam parameters are given in Table~\ref{tab:parameters}. The luminosity biases determined from these inputs by {\tt Guinea-Pig} with leading-order Bhabha events are displayed in Table~\ref{tab:LEPBiases} for $\sqrt{s} = 91.2$\,GeV. The multiplicative correction factor, arising from higher-order effects and energy/angular cuts, as discussed in the previous section, is also indicated. The larger OPAL $k$ factor arises from the tight acollinearity requirement, absent from the selection criteria of the other experiments, which compensates for the reduction due to the initial-state radiation boost.

\begin{table}[htbp]
\begin{center}
\caption{ Luminosity bias determined by {\tt Guinea-Pig} with leading-order Bhabha events, for each of the four LEP experiments in 1993, 1994, and 1995, at $\sqrt{s} = 91.2$\,GeV. The DELPHI entry in 1993 makes use of their first-generation LumiCal, with a narrow acceptance from $55.1$ to $114.6$\,mrad~\cite{Bugge:2629111}. The last line indicates the multiplicative $k$ factor determined from {\tt BHLUMI} with an emulation of the selection criteria of each experiment. \vspace{2mm}}
\label{tab:LEPBiases}
\begin{tabular}{|l|c|c|c|c|}
\hline 
Year        &    ALEPH      & DELPHI       & L3           & OPAL         \\ \hline\hline
 1993       & $-0.0850\%$   & $-0.0389\%$  & $-0.0795\%$  & $-0.0829\%$  \\ \hline
 1994       & $-0.1092\%$   & $-0.0602\%$  & $-0.1018\%$  & $-0.1059\%$  \\ \hline
 1995       & $-0.0832\%$   & $-0.0469\%$  & $-0.0779\%$  & $-0.0809\%$  \\ \hline\hline
 $k$ factor & $1.032$       & $1.033$      & $1.026$      & $1.051$      \\ \hline
\end{tabular}
\end{center}
\end{table}
The luminosity bias is proportional to the bunch population $N$ and is found to scale with the horizontal bunch size like $\sigma_x^{-0.8}$, which explains the year-to-year variation. The dependence on the vertical bunch size $\sigma_y$ and the bunch length $\sigma_z$  is much milder, and amounts to about $\mp 1\%$ ($\mp 6\%$) when varying $\sigma_y$ ($\sigma_z$) by $\pm 40\%$. The luminosity bias is also inversely proportional to the beam energy~\cite{Voutsinas2019}, as more energetic charged particles get less deflected by a given electromagnetic force.
Each experiment collected data at the Z peak ($\langle \sqrt{s} \rangle = 91.224$\,GeV) every year, but also off-peak data ($\langle \sqrt{s} \rangle = 89.446$ and $93.003$\,GeV) in 1993 and 1995, when the luminosity bias was smaller.  The luminosity biases at each centre-of-mass energy, averaged over the three years, are listed in Table~\ref{tab:LEPEnergyBiases}.
\begin{table}[htbp]
\begin{center}
\caption{ Luminosity bias determined for each of the four LEP experiments at 89.4, 91.2, and 93.0\,GeV. The error-weighted averages over the 1993, 1994, and 1995 periods account for the statistical and year-to-year uncorrelated systematic uncertainties~\cite{Abbiendi:1999zx,Barate:1999ce,Abreu:2000mh,Acciarri:2000ai} on the luminosity measurements, as well as for the $k$ factors of Table~\ref{tab:LEPBiases}. 
The last column indicates the luminosity bias averaged over the four experiments at each energy, making use of the total 
experimental (statistical + systematic) and uncorrelated theoretical uncertainties. \vspace{2mm}}
\label{tab:LEPEnergyBiases}
\begin{tabular}{|l|c|c|c|c||c|}
\hline 
{\small $\langle \sqrt{s} \rangle$ (GeV)}  &   ALEPH    & DELPHI     & L3         & OPAL       & LEP      \\ \hline\hline
89.446            & $-0.0889\%$ & $-0.0483\%$ & $-0.0823\%$ & $-0.0877\%$ & $-0.0809\%$ \\ \hline
91.224            & $-0.1029\%$ & $-0.0561\%$ & $-0.0956\%$ & $-0.1064\%$ & $-0.0969\%$ \\ \hline
93.003            & $-0.0855\%$ & $-0.0463\%$ & $-0.0791\%$ & $-0.0844\%$ & $-0.0781\%$ \\ \hline
\end{tabular}
\end{center}
\end{table}

The number of light neutrino species was measured by each LEP experiment to be
\begin{eqnarray}
{\rm ALEPH} & : & N_\nu = 2.983 \pm 0.013,~\text{\cite{Barate:1999ce}} \\
{\rm DELPHI} & : & N_\nu = 2.984 \pm 0.017,~\text{\cite{Abreu:2000mh}} \\
{\rm L3} & : & N_\nu = 2.978 \pm 0.014,~\text{\cite{Acciarri:2000ai}} \\ 
{\rm OPAL} & : & N_\nu = 2.984 \pm 0.013,~\text{\cite{Abbiendi:2000hu}}
\end{eqnarray}
including a fully correlated uncertainty of $\pm 0.0050$ that comes from the common Bhabha cross-section theory error ($\pm 0.0046$), the uncertainty on the QED corrections to the Z lineshape ($\pm 0.0016$), and the uncertainty on $(\Gamma_{\nu\nu}/\Gamma_{\ell\ell})_{\rm SM}$ ($\pm 0.0013$)~\cite{ALEPH:2005ab}. An underestimation of the luminosity at the peak leads to an overestimation of  $\sigma_{\rm {had}}^0$ and, therefore, to an underestimation of $N_\nu$ (Eq.~\ref{eq:NnuForLep}), given by:\footnote{This relation is rounded in Ref.~\cite{ALEPH:2005ab} to $\delta N_{\nu} \simeq - 7.5 \frac{\Delta L}{L}$}
\begin{equation}
  \delta N_{\nu} \simeq - (7.465 \pm 0.005) \times \left. \frac{\Delta L}{L} \right\vert_{91.2\,{\rm GeV}} .
\end{equation}

The luminosity biases in Table~\ref{tab:LEPEnergyBiases} thus result in an increase of $N_\nu$ by $+0.00768$ in ALEPH,  $+0.00419$ in DELPHI, $+0.00713$ in L3, and $+0.00794$ in OPAL, and yield an overall increase $\delta N_\nu = +0.00724$ of the LEP average, with respect to Eq.~\ref{eq:Nnu}. It was checked (Section~\ref{sec:Systematic}) that the data recorded in the 1990-1992 period has a negligible impact on this estimate, because of the very significantly larger luminosity uncertainties (by a factor of 5 to 10) in the early LEP period~\cite{Pietrzyk:1994rk}. 
On the other hand, the most up-to-date calculation of higher-order corrections and the most recent measurements of the Higgs boson and top quark masses yield a small change in $(\Gamma_{\nu\nu}/\Gamma_{\ell\ell})_{\rm SM}$ from $1.99125 \pm 0.00083$~\cite{ALEPH:2005ab} to $1.99083 \pm 0.00025$~\cite{Tanabashi:2018oca}. When this change is included, the correction to $N_\nu$ increases by $+0.00063$ to $\delta N_\nu = +0.00787$, and $N_\nu$ benefits from a small uncertainty reduction. Altogether, the combination of the four LEP experiments for the number of light neutrino species becomes
\begin{equation}
\label{eq:Nnew}
    N_\nu = 2.9919 \pm 0.0081.
\end{equation}

Equivalently, the luminosity biases in Table~\ref{tab:LEPEnergyBiases} result in a reduction of the peak hadronic cross section, $\sigma_{\rm had}^0$, by $-40$\,pb, from $41.540 \pm 0.037$\,nb to $41.500 \pm 0.037$\,nb. It is also interesting to note that the smaller luminosity bias for off-peak data ($-0.0809\%$ and $-0.0781\%$) than for on-peak data ($-0.0969\%$) causes the Z total decay width to slightly increase by $+0.3$\,MeV, from $2.4952 \pm 0.0023$\,GeV~\cite{ALEPH:2005ab} to $2.4955 \pm 0.0023$\,GeV. The Z mass is insignificantly modified by $+22$\,keV. The correlations between the Z mass, the Z width, and the peak hadronic cross section remain untouched with respect to those given in Ref.~\cite{ALEPH:2005ab}. Other electroweak precision observables (asymmetries, ratios of branching fractions) are not affected.

Crosscheck measurements of the {\tt Guinea-Pig} calculations are in principle possible, by exploiting the focusing properties shown in Fig.~\ref{fig:dtheta_vs_phi_and_field} (left) and Fig.~\ref{fig:deltaTheta} (right). Unfortunately, the LEP statistics do not suffice to allow a data-driven determination of the luminosity bias. It would potentially be possible to see some evidence of the focusing effect by observing the $\phi$ modulation of the Bhabha counting rate resulting from the behaviour shown in Fig.~\ref{fig:dtheta_vs_phi_and_field}. If the four experiments are combined, the significance of such a measurement would be about $0.8\sigma$, assuming that potential misalignments of the luminometer system with respect to the interaction point can be corrected for.

It is also possible to define an asymmetry directly proportional to the luminosity bias. To do so, the sample of events selected for the luminosity measurement is split in two sub-samples, according to the sign of the $z_{\rm{vtx}}$ of the events.\footnote{The $z_{\rm{vtx}}$ of Bhabha events can be determined at LEP with a resolution of about $6$~mm, from the intersection of the line joining the two clusters and the $z$ axis in the $(r,z)$ plane.} Four counts $N_{\pm, \pm}$ are defined, representing the number of $\rm e^{\pm}$ (first subscript) measured in the narrow acceptance of one arm of the  LumiCal in events with positive or negative $z_{\rm{vtx}}$ (second subscript). The geometrical change in acceptance induced by the different $z_{\rm{vtx}}$ selections can be corrected for on average. As shown in Fig.~\ref{fig:deltaTheta} (right), since the electrons with $z_{\rm{vtx}} >0$ are more deflected than electrons with $z_{\rm{vtx}} < 0$, $N_{-, +} < N_{-, -}$, such that the asymmetry between these two numbers is proportional to the luminosity bias. The average between the asymmetries built from the electron counts $N_{-, \pm}$ and the positron counts $N_{+, \pm}$ would additionally allow  misalignment effects to partially cancel. This asymmetry, which amounts to about $0.03 \%$, can be observed with a significance of $1.4 \sigma$ using the statistics collected by the four experiments. 

At a future linear $\rm e^+ \rm e^-$ collider, a precise measurement of this asymmetry may be possible, thereby offering an experimental cross-check of the luminosity bias determined by the calculations. Ways to determine the bias at the future circular collider exploit the crossing angle with which the bunches collide, and are described in Ref.~\cite{Voutsinas2019}.

\section{Systematic studies}
\label{sec:Systematic}


A number of simplifying assumptions are made in the previous sections to derive the result presented in Eq.~\ref{eq:Nnew}. In Table~\ref{tab:parameters}, the average luminosity-weighted number of particles per bunch $N$ is inferred from measurements of bunch currents and instantaneous luminosities performed every 15 minutes and recorded in a private database~\cite{HelmutPrivate}. This number was cross-checked
to agree within a few per mil with an analytical calculation involving the mean value of the bunch current distributions in collisions~\cite{Burkhardt:Cham95mb,Burkhardt:306402}, the average coast duration $T$~\cite{Arduini:1996pp} and the average luminosity lifetime $\tau$.\footnote{The average luminosity lifetime is obtained by requiring the average bunch currents of Refs.~\cite{Burkhardt:Cham95mb,Burkhardt:306402} to coincide every year with those computed from the average bunch currents measured at the beginning of the coasts~\cite{Arduini:1996pp}. The luminosity lifetime estimates ($\tau = 16.7$\,hours in 1993, 15\,hours in 1994, and 18\,hours in 1995) are well-compatible with the relation $1/\tau = \xi_y/1\,{\rm hour} + 1/\tau_0$~\cite{Burkhardt:Cham95mb}, with an average vertical beam-beam tune shift $\xi_y$ of 0.027 in 1993, 0.034 in 1994, and 0.023 in 1995~\cite{Arduini:1996pp}, and with $\tau_0 = 30.4$\,hours.} The average horizontal bunch size $\sigma_x$ and bunch length $\sigma_z$ are derived from the variance of the primary vertex position distribution, measured by the experiments~\cite{Abbiendi:1999zx,ALEPHBeamSizes,DELPHIBeamSizes}. 
The $\beta^\ast$ values are taken from Ref.~\cite{Arduini:1996pp}, and the vertical bunch size $\sigma_y$ is obtained by the approximate relation $\sigma_y \sim \sigma_x \times \beta^\ast_y/\beta^\ast_x$~\cite{Brandt:2000xk}. In all instances, it was assumed that all these beam parameters stayed constant over each year. 

The corresponding potential systematic effects were studied as explained below, and are summarized in Table~\ref{tab:SystematicEffects}.

 \begin{itemize}
    \item The bunch current in collisions was measured with an uncertainty of $\pm 2\%$~\cite{HelmutPrivate}, which translates directly to the luminosity bias. 
    \item The average bunch currents for positrons differed from those for electrons by 6\% to 8\%~\cite{Burkhardt:Cham95mb,Burkhardt:306402}, causing a luminosity-bias relative correction of $(-0.6 \pm 0.1)\%$.

    \item The horizontal bunch size and the bunch length agreed among the LEP experiments within 5\%. The luminosity bias varies like $\sigma_x^{-0.8}$, and is therefore uncertain by $\pm 2\%$. The bias variation with $\sigma_z$ is $\pm 0.4\%$.
    \item The vertical bunch size was too small to be measured reliably by the experiments. The relation used above to infer $\sigma_y$ assumes that the horizontal and vertical beam-beam tune shifts $\xi_{x,y}$ were equal~\cite{Burkhardt:261537}. At LEP, this was only approximately the case, with ratios $\xi_{y}/\xi_{x}$ of 1.3 or more~\cite{Brandt:2000xk}. Such a value causes $\sigma_y$ to decrease by $30\%$, and the luminosity bias to relatively increase by $+0.8\%$. An uncertainty of $\pm 0.4\%$ is assigned to this correction. 
    \item The luminosity bias is, to first order, proportional to $N/\sigma_x^{0.8}$. It is determined above from the luminosity-weighted average of $N$ and $\sigma_x$, but a time-dependent analysis would be in order. In the beam-beam limit, the luminosity, the emittances, and the product $\sigma_x \sigma_y$ approximately varied like the beam current~\cite{Burkhardt:261537}, i.e., $\sim \exp(-t/\tau)$. The horizontal and the vertical bunch sizes therefore both varied like $\sim \exp(-t/2\tau)$. The relative effect of a time-dependent analysis can therefore be estimated by comparing the ratio of $\langle \exp(-t/\tau) \rangle$ to $\langle \exp(-t/2\tau) \rangle^{0.8}$ (time-indepen\-dent), to $\langle \exp(-t/\tau)/\exp(-0.4t/\tau) \rangle$ (time-dependent), where $\langle \ \rangle$ means ``luminosity-weighted average''. The time-dependent average is found to be $0.7\%$ smaller. Because the above time dependence of the luminosity and the horizontal  bunch size is only approximate, an uncertainty of half the correction is assigned to this estimate.
    \item The ``technical'' accuracy of {\tt Guinea-Pig} can be evaluated by comparing the {\tt Guinea-Pig} predictions with those of the independent numerical integration shown in Fig.~\ref{fig:dtheta_vs_phi_and_field}. For $\theta_{\rm min} = 31.3$\,mrad, the predictions of the average deflection $\langle \Delta \theta_{\rm FS} \rangle$ agree within $0.4\%$. This comparison yields an uncertainty of $\pm 0.2\%$, to which a statistical uncertainty of $\pm 0.5\%$ is added to account for the size of the {\tt BHLUMI} event samples used to determine the $k$ factors.
    \item The values of $\beta^\ast_x$ and $\beta^\ast_y$ were made to vary by up to $\pm 20\%$, independently at each interaction point, to equalize the luminosities in the four experiments. The luminosity bias is found to be immune to such $\beta^\ast$ changes (other things being equal).
    \item Non-Gaussian beam profiles and partial overlap of colliding bunches could also cause changes in the bias estimate. Regular ``vernier'' scans were performed to adjust the vertical overlap of the beams by varying in steps their vertical separation. The measured resulting beam-beam deflection~\cite{Bovet:306910} allowed the validity of the Gaussian beam profile assumption to be checked, and the vertical overlap $\Delta y$ to be adjusted to better than $0.4\,\mu$m, leading to negligible systematic effects.
    \item The acceptance of each of the LumiCals considered in Table~\ref{tab:AngularRanges} is rounded to the nearest tenth of a mrad, inducing a relative uncertainty of $\pm 0.2\%$ on the luminosity bias.
    \item The statistical, experimental, and theoretical uncertainties on the integrated luminosity, which enter the averaging procedure over all three years and all four experiments, were in general quoted in the LEP experiments' publications in integer units of $10^{-5}$, and sometimes $10^{-4}$, for each LEP running period. The resulting uncertainty on the luminosity bias is evaluated to be $\pm 0.5\%$.
    \item Finally, only the period between 1993 and 1995 has been analysed at this point. The inclusion of the 1991--1992 period~\cite{Barate:1999ce,Abbiendi:2000hu,Bailey:921985,Abreu:259634,Acciarri:261072} causes the average luminosity bias to decrease by a relative $-0.13\%$. The impact of the 1990 data is totally insignificant.    
\end{itemize}
Other effects from, e.g., detailed electromagnetic shower simulation and clustering algorithms applied to final state ${\rm e}^\pm$ prior to applying energy and angular selection criteria, are not easy to estimate precisely, and would require the participation of the LEP experiments. Conservatively, an uncertainty of $\pm 5\%$ is assigned to these effects.

\begin{table}[htbp]
\begin{center}
\caption{Summary of systematic corrections and uncertainties relative to the luminosity bias. Details can be found in the text.\vspace{2mm} }
\label{tab:SystematicEffects}
\begin{tabular}{|l|r|}
\hline 
Source                        &  Systematic effect              \\ \hline\hline
Bunch currents                &  $\pm 2.0\%$                    \\ \hline
${\rm e^+/e^-}$ imbalance     &  $-0.6\%$ $\pm 0.1\%$           \\ \hline
Horizontal bunch size         &  $\pm 2.0\%$                    \\ \hline
Bunch length                  &  $\pm 0.4\%$                    \\ \hline
Vertical bunch size           &  $+0.8\%$ $\pm 0.4\%$           \\ \hline
Time dependence               &  $-0.7\%$ $\pm 0.4\%$           \\ \hline
Technical accuracy            & $\pm 0.6\%$                    \\ \hline
$\beta$ functions at IP       & small                          \\ \hline
Bunch profiles                & small                          \\ \hline
${\rm e^+/e^-}$ bunch overlap & small                          \\ \hline
LumiCal acceptance            & $\pm 0.2\%$                     \\ \hline
Averaging procedure           & $\pm 0.5\%$                     \\ \hline
1990-1992 data                & $-0.1\%$ $\pm 0.0\%$            \\ \hline
Other effects                 &  $\pm 5.0\%$                    \\ \hline\hline
Total                         & $-0.6\%$ $\pm 5.8\%$                    \\ \hline
\end{tabular}
\end{center}
\end{table}

The relative correction on the luminosity bias of $(-0.6 \pm 5.8)\%$ (Table~\ref{tab:SystematicEffects}) yields a small decrease of $\delta N_\nu$ by $(-0.4 \pm 4.1)\times 10^{-4}$, which, everything considered, amounts to $\delta N_\nu = +0.00783 \pm 0.00041$. When this correction is applied, the final LEP combination for the number of light neutrino species becomes
\begin{equation}
    N_\nu = 2.9918 \pm 0.0081,
\end{equation}
while the peak hadronic cross section and the Z width remain as indicated in Section~\ref{sec:neutrinos}.


\section{Conclusions}
\label{sec:summary}

The bias of the luminometer acceptance, induced by the focusing of the final state electrons and positrons from small angle Bhabha scattering by the opposite-charge bunches, has been quantified for the four experiments operating at LEP at and around the Z pole. The  integrated luminosity at the peak has been found to be underestimated by about $0.1 \%$, a bias larger than the uncertainty reported by the experiments in this period. When this bias is corrected for, the number of light neutrino species determined by the combined LEP experiments from the invisible decay width of the ${\rm Z}$ boson increases by $95\%$ of its uncertainty. The corresponding long-standing $2\sigma$ deficit on $N_\nu$ is thereby reduced to about one standard deviation:
\begin{equation*}
    N_\nu = 2.9918 \pm 0.0081. 
\end{equation*}
The luminosity biases at and off the Z peak have also been found to modify the hadronic cross section at the Z peak and the Z width, which become:
\begin{eqnarray*}
\sigma_{\rm had}^0 &=& 41.500 \pm 0.037\,{\rm nb}, \\
\Gamma_{\rm Z} &=& 2.4955 \pm 0.0023\,{\rm GeV}.
\end{eqnarray*}
No other electroweak precision observable is affected. This result has been obtained from averaged LEP operation parameters for each year of the 1990--1995 period. The effects of using averaged values rather than carrying out a time-dependent analysis have been evaluated to be negligible at the level of the current accuracy on the number of neutrino species. Measurements of selected LEP operation parameters have been performed every 15 minutes between 1989 and 2000. These measurements were partially recorded in a private database that still exists, which opens the possibility of a time-dependent analysis. 

\subsection*{Acknowledgements}
We are grateful to Daniel Schulte, Helmut Burkhardt, Gianluigi Arduini, Nicola Bacchetta, Konrad Elsener, Dima El Khechen, Mike Koratzinos, Katsunobu Oide, Dmitry Shatilov, and J\"org Wenninger, for  very useful discussions, suggestions and input that they have brought into this work.


\bibliographystyle{jhep}
\bibliography{biblio}

\end{document}